\documentclass[conference]{IEEEtran}

\usepackage{cite}
\usepackage{amsmath,amssymb,amsfonts}
\usepackage{algorithm,algpseudocode,algorithmicx}
\usepackage{graphicx}
\usepackage{textcomp}
\usepackage{xcolor}
\usepackage{tikz}

\usepackage{bm} 
\usepackage{booktabs} 
\usepackage{multirow} 
\usepackage{stmaryrd} 
\usepackage{enumitem} 

\usepackage{soul}  
\sethlcolor{yellow} 
\definecolor{lightyellow}{RGB}{255, 255, 197} 
\definecolor{lightred}{RGB}{255, 220, 209} 

\usepackage[most]{tcolorbox}
\newtcolorbox{highlightbox}[1]{
    enhanced,
    sharp corners,
    boxrule=0pt,
    colback=#1,        
    top=0pt, bottom=0pt, left=0pt, right=0pt,
    upper left=0pt,
    enlarge top by=-1mm, 
    enlarge bottom by=-1mm,
    boxsep=2pt,
    outer arc=0pt, arc=0pt,
}
\definecolor{algblue}{HTML}{80CED7}
\definecolor{algyellow}{HTML}{FFED66}
\definecolor{algpink}{HTML}{F2ADC1}
\definecolor{alggreen}{HTML}{C5EDAC}

\begin{document}

\title{FPGA Acceleration of Fully Homomorphic Encryption with Adaptive Key Switching}
\author{\IEEEauthorblockN{Zhihan Xu$^{1,*}$, Jayashree Adivarahan$^{1,*}$, Rajgopal Kannan$^{2}$ and Viktor K. Prasanna$^{1}$}
 \IEEEauthorblockA{
$^1$University of Southern California, USA; $^2$ DEVCOM Army Research Office, USA
\\
Email: \{\mbox{zhihanxu, adivarah, prasanna\}@usc.edu}, rajgopal.kannan.civ@army.mil}}
\maketitle

\begingroup\def\thefootnote{*}\footnotetext{~Both authors contributed equally.}\endgroup

\begin{abstract}
Fully Homomorphic Encryption (FHE) enables privacy-preserving cloud services but incurs substantial computation overhead, making hardware acceleration essential. Among FHE operations, key-switching is a major performance bottleneck. Recent cryptographic advances introduce a novel key-switching method (i.e., KLSS) that reduces certain operational complexity but demands higher computational precision than the traditional Hybrid Key Switching (HKS) method. This trade-off leads to distinct computation and memory requirements, making the relative latency of KLSS and HKS highly dependent on hardware parallelism, FHE security parameters, and available on-chip memory capacity, particularly on FPGA platforms, where memory resources and parallelism must be carefully balanced.

In this work, we first propose a memory-efficient KLSS datapath that eliminates off-chip ciphertext transfers. We then develop a performance model to analyze and compare the overheads of both KLSS and HKS. Our analysis reveals that an adaptive solution supporting both methods can achieve lower overall latency than a static method during FHE computation. Guided by the performance model, we design an adaptive FPGA-based FHE accelerator that dynamically selects between HKS and KLSS during computation. We implement the accelerator on an Alveo U280 and evaluate it across multiple FHE benchmarks. Experimental results demonstrate that our adaptive solution achieves a 1.84–3.31$\times$ speedup in bootstrapping latency and a 1.66–2.52$\times$ speedup in secure image classification compared to state-of-the-art FPGA accelerators.


\end{abstract}


\section{Introduction}~\label{sec:intro}
Fully Homomorphic Encryption (FHE) enables computation directly on encrypted data without decryption, providing a powerful foundation for privacy-sensitive cloud applications. Despite its strong security guarantees, FHE introduces enormous computation and memory overheads, often remaining several orders of magnitude slower than computation on plaintext~\cite{de2021does}. As a result, hardware acceleration has become essential for making FHE practical in real-world deployments.


FPGAs have become a popular solution for FHE acceleration. Modern FPGA platforms integrate High Bandwidth Memory (HBM), which helps address the memory bandwidth bottleneck caused by ciphertext expansion, where ciphertexts are typically two to three orders of magnitude larger than plaintexts. Prior FPGA-based FHE accelerators have primarily focused on optimizing individual operations (e.g., key-switching)~\cite{turan2020heaws,yang2023fpga,lu2024fpga,neda2024ciflow} or a specific subroutine (e.g., the Number Theoretic Transform, or NTT)~\cite{kumarathunga2025autontt, singh2025ntt,zhu2026efficient}. More recent work has begun to explore full-system FPGA accelerators~\cite{agrawal2023fab,yang2023poseidon,huang2025effact,xu2026hera} that support complete FHE workloads, including bootstrapping and end-to-end applications.

Among FHE operations, key-switching is one of the most computationally expensive. It can account for a 80\% of total runtime~\cite{fan2025fast}, because it is required for each ciphertext multiplication and rotation operation. Since many FHE benchmarks, such as bootstrapping and practical applications, involve a large number of multiplications and rotations, key-switching is a major performance bottleneck. Consequently, improving the efficiency of key-switching is critical to improving the overall FHE performance.

Recent algorithmic advances have proposed efficient key-switching methods~\cite{bajard2016full,fullrns,halevi2019improved,han2020better,kim2023accelerating}. In particular, the KLSS  method~\cite{kim2023accelerating} reduces the computational complexity by lowering the number of required NTTs, which are the dominant cost in key-switching. However, this reduction in computation comes at the cost of increased memory footprint for evaluation keys and higher precision requirements. Therefore, the practical performance of KLSS depends heavily on the hardware architecture and memory capacity; it is not clear whether KLSS is always beneficial in FPGA-based FHE systems.

In this work, we investigate the efficient implementation of the KLSS key-switching algorithm on FPGA and compare its practical performance with the traditional hybrid key-switching (HKS) method. We determine the practical trade-offs between these two methods and determine conditions when each method is preferable. To the best of our knowledge, this work presents the first FPGA-based FHE compute accelerator that supports the KLSS key-switching method as part of a full FHE pipeline. We show that neither KLSS nor HKS is universally optimal during computation. Therefore, we propose an adaptive architecture that dynamically selects between KLSS and HKS at runtime based on the computation depth and security parameter settings, achieving better overall performance.
The main contributions are as follows:
\begin{itemize}[leftmargin=*]
    \item We propose a memory-efficient datapath for the KLSS tailored to FPGA devices with limited on-chip memory, avoiding intermediate ciphertext transfers to off-chip memory and improving data reuse and arithmetic intensity.
    \item We develop a performance model to analyze the trade-offs between HKS and KLSS. The model captures both computation and memory footprint and shows that neither method is universally optimal during FHE evaluation.
    \item Guided by the performance model, we design an accelerator that supports both HKS and KLSS and dynamically switches between them during runtime based on the computation depth and parameter configuration.
    \item We implement the proposed accelerator on an Alveo U280 and evaluate it across multiple parameter sets and FHE benchmarks. The results show that our adaptive solution outperforms state-of-the-art (SOTA) FPGA-based designs by 1.84–3.31$\times$ on bootstrapping and 1.66–2.52$\times$ on secure image classification.
\end{itemize}

\section{Background}
Among various FHE schemes, we focus on CKKS~\cite{cheon2017homomorphic}, which supports approximate arithmetic over fixed-point real (or complex) numbers and is used in privacy-preserving applications with precision requirements, e.g., machine learning.

\subsection{RNS-CKKS FHE scheme}
Before encryption, sensitive data is packed into a message vector. In CKKS, the message vector $\mathbf{m} \in \mathbb{C}^{N/2}$ is scaled by $\Delta$ and encoded into a plaintext polynomial (Pt), which is then encrypted into a ciphertext (Ct), denoted as $\llbracket \mathbf{m} \rrbracket := (\mathbf{c}_0, \mathbf{c}_1)$. Both Pt and Ct are polynomials of degree $N-1$. Let $\mathcal{R} = \mathbb{Z}[X]/(X^N+1)$; a freshly encrypted Ct resides in $\mathcal{R}_Q^2$, i.e., a pair of polynomials with coefficients modulo $Q$.

The full modulus $Q$ is typically hundreds to thousands of bits, so the Residue Number System (RNS) is used to decompose $Q_L$ into a product of coprime moduli $Q_L=\prod_{i=0}^{L} q_i$, where each $q_i$ is a machine-word-sized prime. Under RNS, each ciphertext polynomial is represented by $L+1$ limbs, i.e., $[\mathbf{c}_0]_{Q_L} \mapsto [\mathbf{c}_0]_{q_0}, \ldots, [\mathbf{c}_0]_{q_L}$. A freshly encrypted ciphertext therefore contains $2(L+1)$ limbs. This variant is referred to as RNS-CKKS~\cite{fullrns}. The parameters used in this work are summarized in Table~\ref{tab:ckks_param}, and we follow the notation in~\cite{xu2026hera}, extended to include KLSS switching.

\subsubsection{CKKS Operations} Assume a plaintext (i.e., Pt') encodes a message vector $\mathbf{m}'$. CKKS mainly supports the following operations based on modular arithmetic on integer polynomials.

\begin{itemize}[left=0pt]

   \item $\mathsf{PAdd}(\llbracket\mathbf{m}\rrbracket, \text{Pt'})\rightarrow\llbracket\mathbf{m}+\mathbf{m'}\rrbracket=(\mathbf{c}_0+\text{Pt'},\mathbf{c}_1)$
   
 \item $\mathsf{PMult}(\llbracket\mathbf{m}\rrbracket, \text{Pt'})\rightarrow\llbracket\mathbf{m}\odot\mathbf{m'}\rrbracket=(\mathbf{c}_0*\text{Pt'},\mathbf{c}_1*\text{Pt'})$

  \item $\mathsf{Add}(\llbracket\mathbf{m}\rrbracket,\llbracket\mathbf{m}'\rrbracket)\rightarrow\llbracket\mathbf{m}+\mathbf{m}'\rrbracket=(\mathbf{c}_0+\mathbf{c}_0',\mathbf{c}_1+\mathbf{c}_1')$  
 \item $\mathsf{Mult}(\llbracket\mathbf{m}\rrbracket,\llbracket\mathbf{m}'\rrbracket)\rightarrow\llbracket\mathbf{m}\odot\mathbf{m}'\rrbracket=(\mathbf{c}_0 * \mathbf{c}_0',\mathbf{c}_0 * \mathbf{c}_1'+\mathbf{c}_1 * \mathbf{c}_0')+\mathsf{KeySwitch}(\mathbf{c}_1 * \mathbf{c}_1', \textbf{evk}_{\text{mult}})$
 
 \item $\mathsf{Rotate}(\llbracket\mathbf{m}\rrbracket,d)\small\rightarrow\llbracket \mathbf{m}\ll d\rrbracket=\mathsf{KeySwitch} (\psi_d(\mathbf{c}_1), \textbf{evk}_{\text{rot}}^{(d)}) \\ +(\psi_d(\mathbf{c}_0),\mathbf{0})$, where $\psi_d$ denotes the automorphism on a polynomial, corresponding to a permutation of coefficients defined as $\psi_d(i) = i \cdot 5^d \bmod N$, for $i \in [0, N)$. The coefficient at index $i$ goes to the new position at $\psi_d(i)$.

\end{itemize}

\begin{table}[t]
\centering
\caption{RNS-CKKS parameters with KLSS-related ones}\label{tab:ckks_param}
\vspace{-0.2cm}
\begin{tabular}{l|l}
    \toprule
    \textbf{Param.} & \textbf{Description} \\ 
    \midrule
    $N$ & \# of coefficients in a Ct/Pt polynomial; a power-of-2\\
    $L$ & Maximum multiplicative level of Ct \\
    $\ell$ & Current remaining level of Ct \\
    $Q$ or $Q_L$ & $=\prod_{i=0}^{L}q_i$, initial modulus of Ct\\
    $Q_\ell$ &  $=\prod_{i=0}^{\ell}q_i$, current modulus of Ct at level $\ell$\\
    \midrule
    $P$ & $=\prod_{i=0}^{k-1}p_i$, auxiliary modulus for HKS\\
    $k$ & \# of moduli in $P$, auxiliary levels for HKS \\
    \textit{dnum} & KeySwitch decomposition number; Max \# of digits \\
    $\alpha$ & $=\lceil(L+1)/dnum\rceil$, \# of limbs in one digit \\
    $\beta$ & $=\lceil (\ell+1)/\alpha\rceil$, \# of digits in Ct during KeySwitch \\
    \midrule
    $T$ & $=\prod_{i=0}^{\alpha'-1}t_i$, special modulus for KLSS \\
    $\alpha'$ & \# of moduli in $T$, used for KLSS Inner Product\\ 
    $\gamma$ & KLSS decomposition length, a fixed value\\
    
    
    $\tilde{\beta}$ & $=\lceil(l+\alpha+1)/\gamma\rceil$, KLSS digits of evaluation keys\\  

    $v$ & Decomposed word length, second decomposition stage KLSS\\  
    
    \midrule
    $L_{\text{boot}}$ & \# of levels consumed by bootstrapping\\
    $L_{\text{eff}}$ &$=L-L_{\text{boot}}$, achievable level after bootstrapping\\
    \bottomrule
\end{tabular}
\vspace{-0.5cm}
\end{table}

\subsubsection{Ct Maintenance Operations}
There are three Ct maintenance operations required in FHE: key-switching ($\mathsf{KeySwitch}$), rescaling ($\mathsf{Rescale}$), and bootstrapping ($\mathsf{Bootstrap}$).

\textit{Key-switching:} Key-switching is required after each $\mathsf{Mult}$ and $\mathsf{Rotate}$. It consists primarily of an inner product (i.e., $\mathsf{KeyInnerProd}$) between the input polynomial and the evaluation key (\textbf{evk}). The size and structure of the evaluation key differ between the KLSS and HKS methods. We describe the KLSS and HKS algorithms in Sec.~\ref{sec:key-switch} and provide a detailed complexity analysis in Sec.~\ref{sec:complex}.

\textit{Rescaling:} Before encryption, the message is scaled by $\Delta$, which grows to $\Delta^2$ after each multiplication. Rescaling restores the scale by mapping $\llbracket \mathbf{m} \rrbracket \in \mathcal{R}_{Q_\ell}$ to $\llbracket q_\ell^{-1} \odot \mathbf{m} \rrbracket \in \mathcal{R}_{Q_{\ell-1}}$~\cite{fullrns}, implemented as $(\mathbf{c}[i]-\mathbf{c}[\ell])q_\ell^{-1} \bmod q_i$. It removes the last modulus $q_\ell$, consuming one multiplicative level.

\textit{Bootstrapping:} 
When the Ct level reaches 0, no further multiplications are possible, and bootstrapping is required to refresh the levels. Bootstrapping enables deep evaluation circuits without returning ciphertexts to the client. However, it involves many $\mathsf{Mult}$ and $\mathsf{Rotate}$, making key-switching the dominant cost. Since $\mathsf{Bootstrap}$ itself consumes levels ($L_{boot}$), only $L_{eff}$ levels can be recovered instead of the full $L$.

\subsection{Key-switching}\label{sec:key-switch}
Key-switching involves several FHE subroutines, including NTT ($\mathsf{NTT}$) and its inverse ($\mathsf{iNTT}$), RNS basis conversion ($\mathsf{BConv}$), and $\mathsf{KeyInnerProd}$. 
The NTT is the finite-field analogue of the Fast Fourier Transform~\cite{liang2022number} and is used to accelerate polynomial multiplication by converting polynomials into the evaluation domain, where multiplication can be performed element-wise. As a result, Ct and Pt polynomials are typically stored in the evaluation domain.

However, $\mathsf{BConv}$ (e.g., $Q\mapsto P$) cannot be performed in the evaluation domain. Therefore, an $\mathsf{iNTT}$ is required before $\mathsf{BConv}$, followed by an $\mathsf{NTT}$ to convert the result back to the evaluation domain. $\mathsf{BConv}$ is performed by computing inner products across limbs in the original modulus basis and reducing the results modulo the new basis. Consequently, extending one limb requires one inner product across all limbs in the original basis. $\mathsf{ModUp}$ and $\mathsf{ModDown}$ denote $\mathsf{BConv}$ that increases and decreases the modulus, respectively.

In HKS, the \textbf{evk} is a $2 \times \beta$ matrix of polynomials, where each polynomial has modulus $PQ_\ell$ with $\ell+k+1$ limbs. To perform $\mathsf{KeyInnerProd}$, the input Ct polynomial is first decomposed into $\beta$ digits, each containing $\alpha$ limbs. $\mathsf{ModUp}$ is then applied to extend each digit from modulus $Q_{\alpha-1}$ to $PQ_\ell$. The digit vector is multiplied with the \textbf{evk} matrix to perform $\mathsf{KeyInnerProd}$, producing two polynomials in modulus $PQ_\ell$. Finally, $\mathsf{ModDown}$ reduces the modulus of each polynomial back to $Q_\ell$ with $\ell+1$ limbs. Algorithm~\ref{algo:hks} summarizes HKS.

\begin{algorithm}[t]
\caption{$\mathsf{KeySwitch_{HKS}}(\mathbf{c}_{in}, \textbf{evk})\rightarrow(\mathbf{c}_0, \mathbf{c}_1)$}
\label{algo:hks}
\begin{flushleft}
\textbf{Input:} $\mathbf{c}_{in} \in \mathcal{R}_{Q_\ell}$, $\textbf{evk} \in \mathcal{R}_{PQ_\ell}^{2 \times \beta}$ \\
\textbf{Output:} $(\mathbf{c}_0, \mathbf{c}_1)\in\mathcal{R}_{Q_\ell}^2$
\end{flushleft}
\vspace{-0.2cm}
\begin{algorithmic}[1]
\State{$(\mathbf{d}_j)_ {j\in[0,\beta)}\leftarrow\mathbf{c}_{in}$}
\algorithmiccomment{Decompose into $\beta$ digits}
\begin{tcolorbox}[colback=algpink, standard jigsaw, opacityback=0.6, boxrule=0pt, sharp corners, boxsep=1pt, left=2pt, right=2pt, top=0pt, bottom=0pt]
\For{$j=0$ to $\beta-1$}
    \State $\mathbf{d}_j'\leftarrow\mathsf{iNTT}(\mathbf{d}_j)$\algorithmiccomment{$\alpha$ limbs}
    \State $\hat{\mathbf{d}}_j'=\{\text{d}_j^{\alpha},\cdots,\text{d}_j^{\ell+k}\}\leftarrow\mathsf{BConv(\mathbf{d}_j')}$
    \State $\hat{\mathbf{d}}_j\leftarrow\{\mathbf{d}_j,\mathsf{NTT}(\hat{\mathbf{d}}_j')\}$\algorithmiccomment{$\mathcal{R}_{PQ_\ell}$}
\EndFor
\State{\algorithmiccomment{$\mathsf{ModUp}:\mathcal{R}_{Q_{\alpha-1}}\mapsto\mathcal{R}_{PQ_\ell}$}}
\end{tcolorbox}

\begin{tcolorbox}[colback=algyellow, standard jigsaw, opacityback=0.6, boxrule=0pt, sharp corners, boxsep=1pt, left=2pt, right=2pt, top=0pt, bottom=0pt]
\State $(\hat{\mathbf{c}}_0, \hat{\mathbf{c}}_1) \gets (0, 0) $\algorithmiccomment{$\mathcal{R}_{PQ_\ell}^2$}
\For{$j = 0$ to $\beta - 1$}
    \State $(\hat{\mathbf{k}}_0, \hat{\mathbf{k}}_1) \gets \textbf{evk}[j]$ \Comment{$j$-th column of $\textbf{evk}$, $\mathcal{R}_{PQ_\ell}^2$}
    \State $(\hat{\mathbf{c}}_0, \hat{\mathbf{c}}_1)  += (\hat{\mathbf{d}}_j \odot \hat{\mathbf{k}}_0, \hat{\mathbf{d}}_j \odot \hat{\mathbf{k}}_1)$
\EndFor
\State\algorithmiccomment{$\mathsf{KeyInnerProd}$ in $\mathcal{R}_{PQ_\ell}$}
\end{tcolorbox}
\begin{tcolorbox}[colback=algblue, standard jigsaw, opacityback=0.6, boxrule=0pt, sharp corners, boxsep=1pt, left=2pt, right=2pt, top=0pt, bottom=0pt]
\For{$i=0$ to $1$}
    \State $\hat{\mathbf{c}}_i'\leftarrow\mathsf{iNTT}(\hat{\mathbf{c}}_i[0, k-1])$\algorithmiccomment{$\mathcal{R}_{P}$}
    \State $\mathbf{c}_i'=\{\text{c}_i^{k},\cdots,\text{c}_i^{\ell+k}\}\leftarrow\mathsf{BConv(\hat{\mathbf{c}}_i')}$
    \State $\mathbf{c}_i\leftarrow\hat{\mathbf{c}}_i[k, \ell+k]-\mathsf{NTT}(\mathbf{c}_i')$\algorithmiccomment{$\mathcal{R}_{Q_\ell}$}
\EndFor
\State\algorithmiccomment{$\mathsf{ModDown}:\mathcal{R}_{PQ_\ell}\mapsto\mathcal{R}_{Q_\ell}$}
\end{tcolorbox}
\end{algorithmic}
\end{algorithm}

\begin{algorithm}[t!]
\caption{$\mathsf{KeySwitch_{KLSS}}(\mathbf{c}_{in}, \textbf{evk})\rightarrow(\mathbf{c}_0, \mathbf{c}_1)$}
\label{algo:klss}
\begin{flushleft}
\textbf{Input:} $\mathbf{c}_{in} \in \mathcal{R}_{Q_\ell}$, $\textbf{evk} \in \mathcal{R}_{T}^{2 \times \beta \times\tilde{\beta}}$ \\
\textbf{Output:} $(\mathbf{c}_0, \mathbf{c}_1)\in\mathcal{R}_{Q_\ell}^2$
\end{flushleft}
\vspace{-0.2cm}
\begin{algorithmic}[1]
\State{$(\mathbf{d}_j)_ {j\in[0,\beta)}\leftarrow\mathbf{c}_{in}$}
\algorithmiccomment{Decompose into $\beta$ digits}
\begin{tcolorbox}[colback=algpink, standard jigsaw, opacityback=0.6, boxrule=0pt, sharp corners, boxsep=1pt, left=2pt, right=2pt, top=0pt, bottom=0pt]
\For{$j=0$ to $\beta-1$}
    \State $\mathbf{d}_j'\leftarrow\mathsf{iNTT}(\mathbf{d}_j)$\algorithmiccomment{$\alpha$ limbs}
    \State $\hat{\mathbf{d}}_j'=\{\text{d}_j^{0},\cdots,\text{d}_j^{\alpha'-1}\}\leftarrow\mathsf{BConv(\mathbf{d}_j')}$
    \State $\hat{\mathbf{d}}_j\leftarrow\mathsf{NTT}(\hat{\mathbf{d}}_j')$\algorithmiccomment{$\mathcal{R}_{T}$}
\EndFor
\State{\algorithmiccomment{$\mathsf{ModUp}:\mathcal{R}_{Q_{\alpha-1}}\mapsto\mathcal{R}_{T}$}}
\end{tcolorbox}

\begin{tcolorbox}[colback=algyellow, standard jigsaw, opacityback=0.6, boxrule=0pt, sharp corners, boxsep=1pt, left=2pt, right=2pt, top=0pt, bottom=0pt]

\For{$i=0$ to $\tilde{\beta}-1$}\Comment{Double decomposition}
\State $(\hat{\mathbf{c}}_{(0,i)}, \hat{\mathbf{c}}_{(1,i)}) \gets (0, 0) $\algorithmiccomment{$\mathcal{R}_{T}^2$}
\For{$j = 0$ to $\beta - 1$}
    \State $(\hat{\mathbf{k}}_0, \hat{\mathbf{k}}_1) \gets \textbf{evk}[i][j]$ \Comment{$\mathcal{R}_{T}^2$}
    \State $(\hat{\mathbf{c}}_{(0,i)}, \hat{\mathbf{c}}_{(1,i)}) += (\hat{\mathbf{d}}_j \odot \hat{\mathbf{k}}_0,\hat{\mathbf{d}}_j \odot \hat{\mathbf{k}}_1)$
\EndFor
\EndFor
\State\algorithmiccomment{$\mathsf{KeyInnerProd}$ in $\mathcal{R}_{T}$}
\end{tcolorbox}

\begin{tcolorbox}[colback=alggreen, standard jigsaw, opacityback=0.6, boxrule=0pt, sharp corners, boxsep=1pt, left=2pt, right=2pt, top=0pt, bottom=0pt]
\For{$j=0$ to $1$}
\For{$i=0$ to $\tilde{\beta}-1$}
    \State $\bar{\mathbf{c}}_{(j,i)}'\leftarrow\mathsf{BConv}(\mathsf{iNTT}(\hat{\mathbf{c}}_{(j,i)}))$ 
   \State \Comment{$\bar{\mathbf{c}}_{(j,i)}'=\{\bar{\text{c}}_{(j,i)}^{0},\cdots,\bar{\text{c}}_{(j,i)}^{\lceil(\ell+\alpha+1)/\tilde{\beta}\rceil-1}\}$}
\EndFor
\EndFor
\State \Comment{Limb Recovery: $2\alpha'\tilde{\beta}\mapsto 2(\ell+\alpha+1)$ limbs}
\end{tcolorbox}

\begin{tcolorbox}[colback=algblue, standard jigsaw, opacityback=0.6, boxrule=0pt, sharp corners, boxsep=1pt, left=2pt, right=2pt, top=0pt, bottom=0pt]
\For{$j=0$ to $1$}
    \State $\mathbf{c}_i\hspace{-0.1cm}=\hspace{-0.1cm}\{\text{c}_i^{0},\cdots,\text{c}_i^{\ell}\}\tiny{\leftarrow}\mathsf{ModDown}(\{\bar{\mathbf{c}}_{(j,0)}',\cdots, \bar{\mathbf{c}}_{(j,\tilde{\beta}-1)}'\})$
\EndFor
\State\algorithmiccomment{$\mathsf{ModDown}:\ell+\alpha+1 \mapsto\ell+1$ limbs}
\end{tcolorbox}
\end{algorithmic}
\end{algorithm}

KLSS is also known as the double-decomposition method, as the \textbf{evk} is decomposed into a $2 \times \beta \times \tilde{\beta}$ tensor of polynomials, where each polynomial has $\alpha'$ limbs. As a result, the $\mathsf{BConv}$ operation in $\mathsf{ModUp}$ only needs to generate or extend $\alpha'$ new limbs in $\mathcal{R}_T$, which is typically much smaller than the $\ell + k + 1 - \alpha$ limbs required in HKS. Consequently, the number of required $\mathsf{NTT}$ operations is significantly reduced.
Prior KLSS-related work has shown that using a large coefficient modulus (i.e., $\log t_i > 60$ bits) is beneficial, as it leads to a small $\alpha'$ and thus reduces the number of $\mathsf{NTT}$~\cite{kim2023accelerating,fan2025fast,jin2025gpu}. 
During $\mathsf{KeyInnerProd}$, KLSS performs similarly to HKS by multiplying and accumulating along the $\beta$ dimension. After $\mathsf{KeyInnerProd}$, the result is a $2 \times \tilde{\beta}$ matrix of polynomials in $\mathcal{R}_T$. An additional step in KLSS is limb recovery, where the number of limbs per row is recovered to $\ell + \alpha + 1$ limbs using $\mathsf{BConv}$, requiring $\lceil (\ell + \alpha + 1)/\tilde{\beta} \rceil$ extensions per polynomial. The polynomials in each row are then aggregated along the $\tilde{\beta}$ dimension to produce two polynomials with $\ell + \alpha + 1$ limbs. Finally, $\mathsf{ModDown}$ reduces the modulus back to $Q_\ell$ with $\ell + 1$ limbs. Algorithm~\ref{algo:klss} summarizes KLSS.

\section{Memory-Efficient KLSS Datapath on FPGA}
\begin{figure}[t!]
    \centering
    \includegraphics[width=1.0\linewidth]{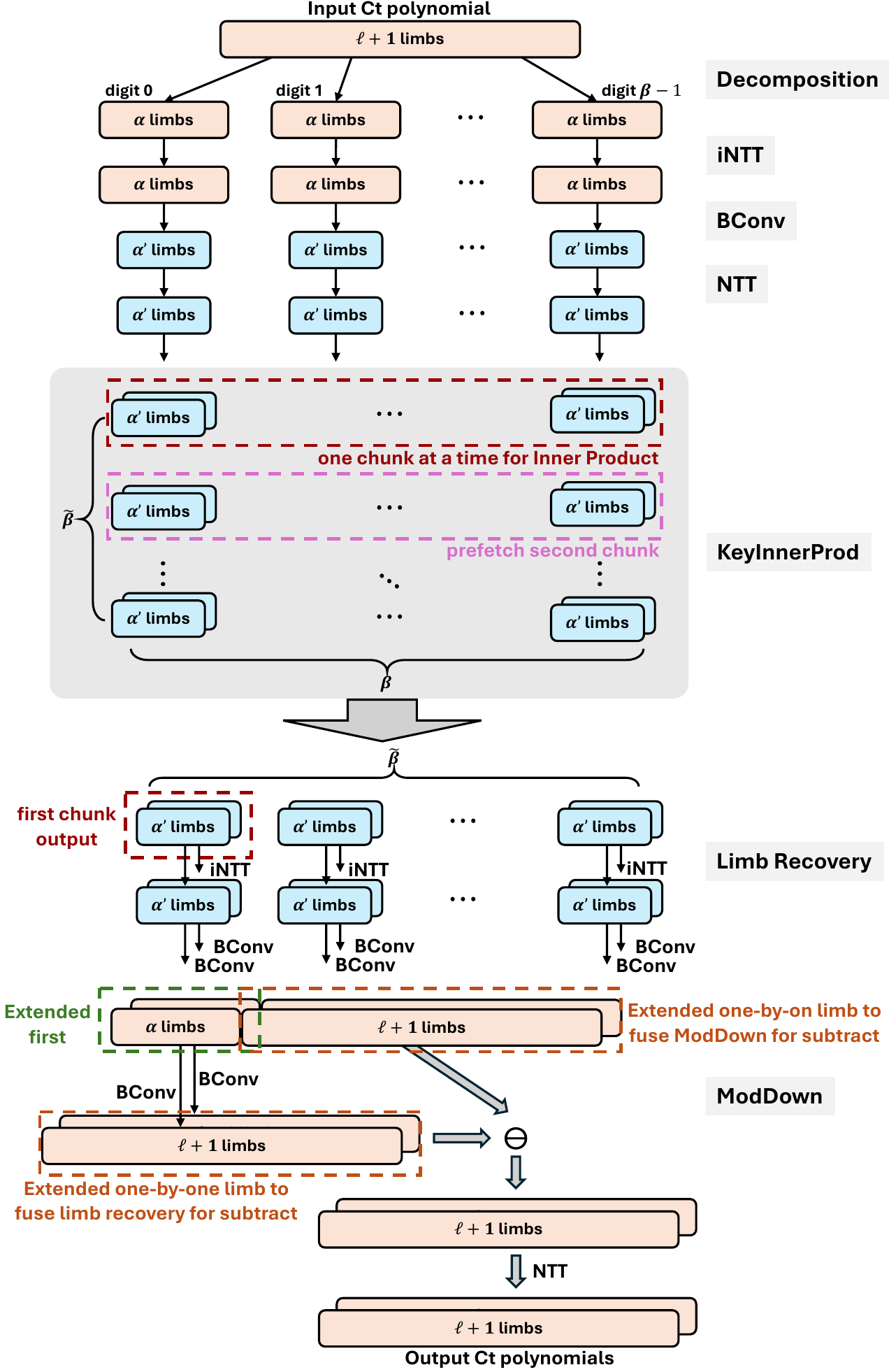} 
    \caption{Proposed memory-efficient KLSS datapath on FPGA}
    \label{fig:klss_datapath}
    \vspace{-0.5cm}
\end{figure}

HKS has been extensively studied in prior work~\cite{agrawal2023fab,xu2026hera} and can be implemented entirely on-chip on FPGA platforms without transferring intermediate ciphertexts to off-chip memory, significantly reducing memory traffic and latency. In contrast, KLSS is a relatively new technique; no prior work has optimized the KLSS datapath for FPGAs. KLSS introduces two main challenges: (1) a decomposed evaluation key (\textbf{evk}) that increases the buffer requirements, and (2) an additional limb recovery step that requires extra basis conversions and generates a large number of intermediate Ct limbs that are difficult to buffer on-chip. To address these challenges, we propose a memory-efficient KLSS datapath (Fig.~\ref{fig:klss_datapath}) that avoids intermediate Ct transfers to off-chip memory on the target FPGA device with 43\,MB SRAM capacity.

Our analysis is based on a practical parameter set that supports bootstrapping and achieves 128-bit security (Set-2 in Table~\ref{tab:parameter_sets}). We first analyze the memory footprint during $\mathsf{ModUp}$. The input Ct polynomial has $\ell+1$ limbs stored on-chip, and after digit decomposition each digit contains $\alpha$ limbs. During $\mathsf{ModUp}$, $\mathsf{BConv}$ generates $\alpha'$ new limbs in $\mathcal{R}_T$, resulting in a memory footprint of $\beta(\alpha+\alpha')$ limbs (13\,MB). The $\alpha'$ limbs in $\mathcal{R}_T$ use a large coefficient modulus (62 bits in our implementation). After generating the $\alpha'$ limbs, the original $\alpha$ limbs can be discarded to free memory.

The next step is $\mathsf{KeyInnerProd}$, which requires accessing the decomposed evaluation keys. The input ($\beta\alpha'$ limbs) and output ($2\alpha'\tilde{\beta}$ limbs) together require about 15\,MB of buffer space. However, KLSS requires $2\beta\tilde{\beta}$ \textbf{evk} polynomials, each consisting of $\alpha'$ limbs, resulting in a total evk size of about 35\,MB, which exceeds the remaining on-chip memory. To address this, we reorganize the \textbf{evk} as a 2-D $\beta \times \tilde{\beta}$ matrix and partition it row-wise into $\tilde{\beta}$ chunks. Each chunk ($2\beta\alpha'$ limbs) is streamed on-chip to perform $\mathsf{KeyInnerProd}$ with the $\beta$ input digits, producing two output digits with $\alpha'$ limbs each. After $\tilde{\beta}$ iterations, all $2\tilde{\beta}$ output digits are generated. We further adopt a double-buffering strategy to overlap \textbf{evk} streaming with computation. This requires buffering two \textbf{evk} chunks simultaneously (12\,MB). Together with the input and output buffers ($\sim$15\,MB), the total memory needed is within the on-chip SRAM capacity, allowing us to avoid intermediate ciphertext transfers to off-chip memory.

An additional step for KLSS is limb recovery. A straightforward implementation would perform $\mathsf{BConv}$ on all $2\tilde{\beta}$ digits simultaneously, requiring buffering both input ($2\tilde{\beta}\alpha'$ limbs) and output ($2(\ell+\alpha+1)$ limbs), totaling about 38\,MB and significantly increasing memory pressure. Instead, we fuse the limb recovery step with $\mathsf{ModDown}$. Rather than recovering all $\ell+\alpha+1$ limbs at once, we first recover the $\alpha$ limbs required for each $\mathsf{ModDown}$, while the remaining $\ell+1$ limbs are generated one-by-one and immediately used for subtraction in $\mathsf{ModDown}$. This limb-by-limb fusion eliminates the need to buffer all recovered limbs simultaneously. In this design, only the limb recovery input ($2\tilde{\beta}\alpha'$ limbs) and the $\mathsf{ModDown}$ input ($2\alpha$ limbs) are buffered, requiring about 19\,MB and fitting within the available SRAM capacity. 

Overall, the proposed datapath enables KLSS to be executed entirely on-chip. By combining \textbf{evk} chunking, double buffering, and fusion of limb recovery with $\mathsf{ModDown}$, the design eliminates intermediate Ct transfers to off-chip memory and reduces memory requirements.

\section{Analytical Model for HKS-KLSS Selection}
In this section, we first summarize the computational complexity of the FHE primitives for both HKS and KLSS. We then develop a performance model that maps algorithmic complexity to execution cycles by incorporating hardware architecture and device characteristics. The model is used to determine whether HKS or KLSS should be applied under different benchmarks and FHE parameter sets.

\subsection{Computational Complexity Analysis}\label{sec:complex}
Table~\ref{tab:complexity} summarizes the computational complexity of the major primitive operations for HKS and KLSS. KLSS significantly reduces the number of $\mathsf{NTT}$, because $\alpha'$ can be small when using a large coefficient modulus in $\mathcal{R}_T$, as discussed in Sec.~\ref{sec:key-switch}. However, KLSS may increase the complexity of other primitives depending on the parameter set. For example, under small parameter sets (e.g., Set-1 in Table~\ref{tab:parameter_sets}), the $\mathsf{BConv}$ complexity of KLSS can be higher than that of HKS. Therefore, a performance model is needed to capture the combined effects of computation and memory behavior and to determine the preferable method under different conditions.

\begin{table}[t]
  \centering
  \caption{Complexity Comparison of HKS and KLSS}
  \vspace{-0.2cm}
  \label{tab:complexity}
  \resizebox{\columnwidth}{!}{%
    \begin{tabular}{ccc}
      \toprule
      \textbf{Primitive} & \textbf{HKS} & \textbf{KLSS} \\
      \midrule
      \midrule
      $\mathsf{iNTT}$ & $\beta\alpha + 2k$ & $\beta\alpha + 2\alpha' \tilde{\beta}$ \\
      \midrule
      $\mathsf{NTT}$ & $\beta(\ell+k+1-\alpha) + 2(\ell+ 1)$ & $\beta\alpha' + 2(\ell+ 1)$ \\
      \midrule
      \multirow{2}{*}{$\mathsf{BConv}$} & $\beta(\alpha(\ell+k+1-\alpha))+$  & $\beta(\alpha\alpha')  + 2(\alpha(\ell+1))+$ \\ 
       & $2(k(\ell+1))$ & $2\tilde{\beta}(\alpha'\lceil(\ell+\alpha+1)/\tilde{\beta}\rceil)$ \\
      \midrule
      $\mathsf{KeyInnerProd}$ & $2\beta (\ell + 1 + k)$ & $2\alpha' \beta\tilde{\beta}$ \\
      \bottomrule
    \end{tabular}%
  }
  \vspace{-0.4cm}
\end{table}



\subsection{Performance Modeling}
Following prior FPGA-based FHE architectures, we assume the accelerator processes $dp$ coefficients per cycle, where $dp$ represents the SIMD parallelism (i.e., the number of compute lanes). The accelerator can perform $dp$ modular multiplications, additions, and multiply-add operations per cycle. We assume permutation operations are also processed at the same throughput, and that on-chip memory can supply $dp$ coefficients per cycle without access conflicts once data is loaded from off-chip memory.

In this work, we use different coefficient sizes for HKS and KLSS. KLSS benefits from a large coefficient modulus in $\mathcal{R}_T$ (we use 62-bit), while other operations use smaller coefficients (e.g., 32-bit), which are sufficient for the precision requirements of most FHE workloads and are commonly used in prior accelerators~\cite{kim2023sharp,xu2025fast,fan2025fast}. Smaller coefficients allow higher SIMD parallelism due to lower DSP usage and also increase the available ciphertext levels, reducing the frequency of bootstrapping. As a result, HKS can achieve higher parallelism than KLSS under the same resource budget. We denote this difference using a scaling factor $\delta$, where $dp_{HKS} = \delta \cdot dp_{KLSS}$.

The total latency of key-switching is modeled as:
\begin{equation}
    \mathcal{T}_{\mathsf{KeySwitch}} = \sum_{i\in OP}\mathcal{T}_{op_i},
\end{equation}
where $OP=\{\mathsf{iNTT, NTT, BConv, KeyInnerProd}\}$. For each primitive operation,
\begin{equation}
    \mathcal{T}_{op_i}=\max(\mathcal{T}_{op_i-comp}, \mathcal{T}_{op_i-mem}),
\end{equation}
where $\mathcal{T}_{op_i-comp}$ and $\mathcal{T}_{op_i-mem}$ denote computation time and memory time, respectively.

The computation time is derived from the algorithmic complexity in Table~\ref{tab:complexity}:
\begin{equation}
    \mathcal{T}_{op_i-comp}=
    \begin{cases}
    \mathcal{C}_{op_i}\cdot\frac{N\log N}{dp} & \text{if } op_i \in \{\mathsf{NTT,iNTT}\} \\
    \mathcal{C}_{op_i}\cdot\frac{N}{dp} & \text{otherwise}
    \end{cases}
\end{equation}
where $\mathcal{C}_{op_i}$ is the computational complexity of primitive $op_i$.

The memory time is modeled as:
\begin{equation}
    \mathcal{T}_{op_i-mem}=F_{clk}\cdot\frac{\mathcal{B}_{op_i-read}+\mathcal{B}_{op_i-write}}{\mathcal{BW}_{max}},
\end{equation}
where $\mathcal{BW}_{max}$ is the peak off-chip memory bandwidth, $\mathcal{B}_{op_i-read}$ and $\mathcal{B}_{op_i-write}$ denote total data transfer volume, and $F_{clk}$ is the clock frequency.

\subsection{Motivating Example}
On the target platform (Alveo U280), we can instantiate 512 32-bit modular multipliers and 256 62-bit modular multipliers, resulting in $dp_{HKS}=512$, $dp_{KLSS}=256$, and $\delta=2$. We assume a peak HBM bandwidth of 460\,GB/s and a target frequency of 300\,MHz. We evaluate practical parameter set (Set-2 in Table~\ref{tab:parameter_sets}) that supports bootstrapping.

\begin{figure}[t]
    \centering
    \includegraphics[width=\linewidth]{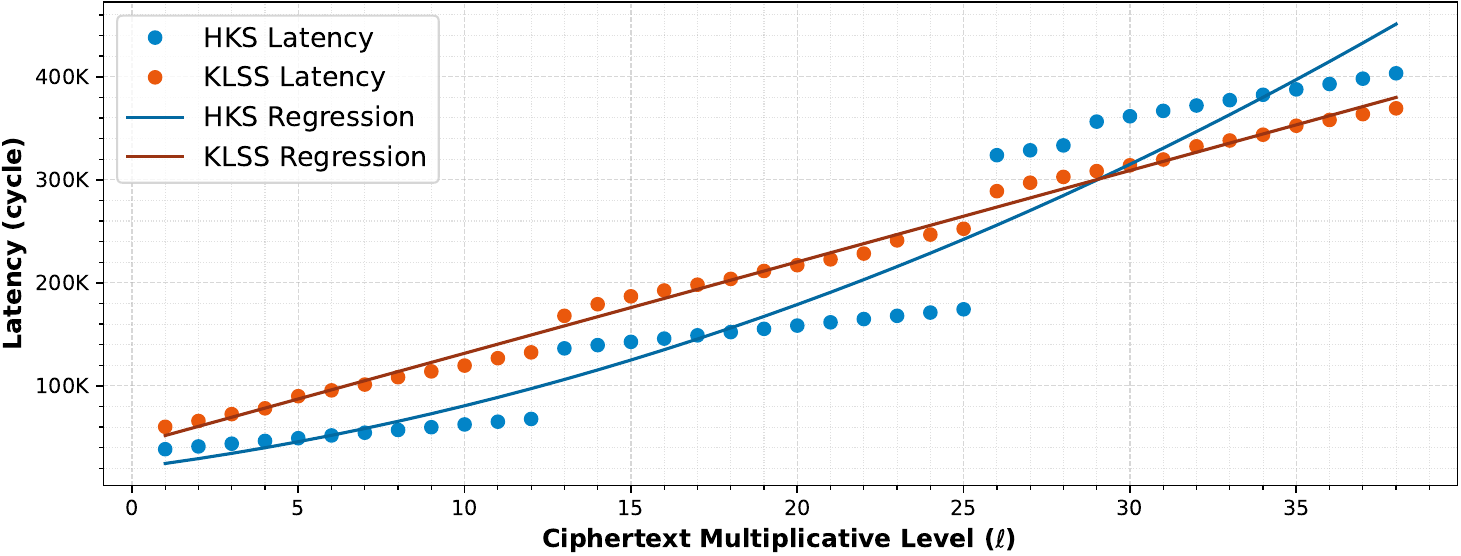}
    \vspace{-0.4cm}
    \caption{Performance trade-off between HKS and KLSS across Ct levels}
    \label{fig:motivation}
\end{figure}


With the proposed memory-efficient KLSS datapath and prior on-chip HKS datapaths~\cite{agrawal2023fab,xu2026hera}, intermediate Ct transfers to off-chip memory are avoided. Therefore, memory traffic mainly consists of input Ct, twiddle factors, and evaluation keys. Fig.~\ref{fig:motivation} shows the modeled latency of HKS and KLSS across Ct levels.
The results show that KLSS outperforms HKS at high Ct levels due to the reduced number of limbs during $\mathsf{ModUp}$, which significantly lowers the cost of $\mathsf{NTT}$, $\mathsf{KeyInnerProd}$, and $\mathsf{BConv}$. However, as the Ct level decreases, the KLSS digit size $\alpha'\tilde{\beta}$ approaches $\ell+k+1$ of HKS, reducing the algorithmic advantage of KLSS. Combined with the higher SIMD parallelism of HKS, HKS becomes more efficient at low Ct levels. This crossover point motivates an adaptive accelerator that supports both KLSS and HKS and dynamically selects the more efficient method during FHE evaluation.

\section{The Proposed Accelerator}

\begin{figure}[t]
    \centering
    \includegraphics[width=\linewidth]{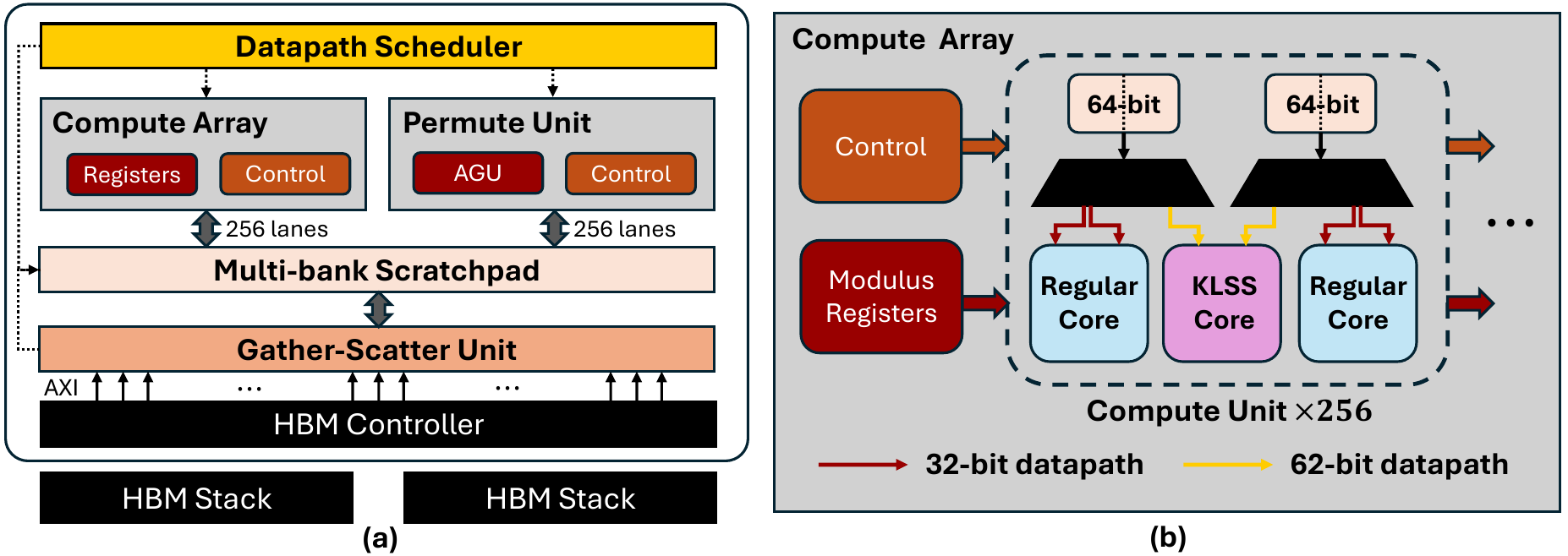}
    \caption{(a) Overall architecture of the accelerator; (b) Architecture of the compute array. Each core supports three-operand instructions (add, multiply, or MAC). Only two inputs are shown for simplicity. The third operand is maintained in an additional 64-bit register.}
    \label{fig:hardware}
    \vspace{-0.3cm}
\end{figure}

The overall architecture of the proposed accelerator with the HBM subsystem is shown in Fig.~\ref{fig:hardware}(a). The accelerator mainly consists of a compute array, a scratchpad, a permute unit, and a datapath scheduler. The architecture supports adaptive parallelism for both KLSS and regular execution by configuring the compute array and permute unit to operate under two parallelism modes. The datapath scheduler orchestrates computation and permutation tasks and determines how to overlap these tasks under different parallelism configurations. In addition, it performs runtime switching between KLSS and HKS based on Ct metadata. The following subsections describe how adaptive computation, permutation, and scheduling are supported in the proposed architecture.

\subsection{Adaptive Compute Array}
Fig.~\ref{fig:hardware}(b) shows the compute array with 256 compute units and a register file storing parameters such as the modulus chain. Each compute unit contains two regular cores and one KLSS core. Each regular core has one 32-bit modular multiplier and adder, while each KLSS core has one 62-bit modular multiplier and adder.
Each core supports configurable operations, including modular multiplication, addition/subtraction, and multiply-add. The multiply-add operation is used in inner-product-based primitives (e.g., $\mathsf{BConv}$ and $\mathsf{KeyInnerProd}$) and butterfly operations in $\mathsf{(i)NTT}$. 

To support adaptive execution, the compute array operates in two parallelism modes. For regular execution over $\mathcal{R}_Q$ with 32-bit coefficients, both regular cores in each compute unit are used, providing an effective parallelism of 512 lanes. For KLSS execution over $\mathcal{R}_T$ with 62-bit coefficients, only the KLSS cores are used, resulting in 256-way parallelism. In this way, the compute array adapts its parallelism based on the arithmetic precision requirements.

The modular multiplier uses a fully pipelined Barrett reduction design. Each 32-bit modular multiplier uses 11 DSPs, resulting in $11 \times 512 = 5632$ DSPs for regular cores. Given the platform DSP budget (9024 DSPs), we adopt an efficient 62-bit modular multiplier design~\cite{kim2019fpga} that uses only 12 DSPs. 
Specifically, since $\alpha'$ is small, we use special moduli (i.e., Solinas primes~\cite{solinas1999generalized}) for $\mathcal{R}_T$. With Solinas primes, the reduction step can be implemented using shift and add operations instead of multipliers~\cite{kim2019fpga}. This optimization is not applicable to $\mathcal{R}_Q$ due to the large number of distinct moduli ($q_i$).

\subsection{Adaptive Permute Unit}\label{sec:perm}
The permutation unit performs coefficient permutations required in $\mathsf{(i)NTT}$ and automorphism. Due to irregular memory access patterns, permutation is a major bottleneck in FHE accelerators~\cite{yang2023poseidon}. We implement the permutation unit using a streaming permutation network~\cite{chen2015automatic}, which is rearrangeable, non-blocking, and supports arbitrary permutation patterns.

\begin{figure}[t]
    \centering
    \includegraphics[width=\linewidth]{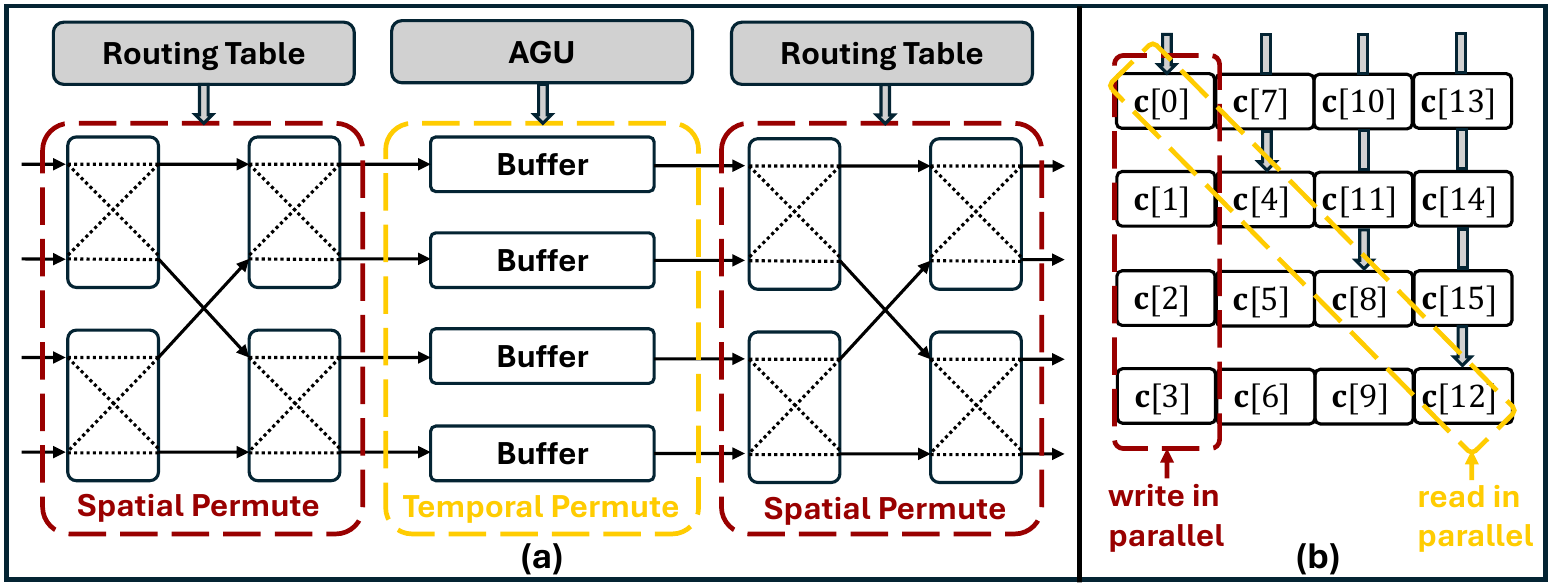}
    \caption{(a) Permute unit with an example of $dp=4$; (b) Temporal permutation example with a stride 4 for 16-point $\mathsf{(i)NTT}$}
    \label{fig:perm}
    \vspace{-0.3cm}
\end{figure}

The permutation unit provides a physical throughput of 256 coefficients per cycle. As shown in Fig.~\ref{fig:perm}(a), the network consists of two spatial permutation networks based on butterfly topology and one temporal permutation network. Each spatial network has $\log dp$ stages composed of $2 \times 2$ switches, while the temporal network uses reorder buffers to permute data across cycles. Routing tables store switch configuration bits, and the address generation unit (AGU) generates memory addresses for temporal permutations, including $\mathsf{(i)NTT}$ permutations based on initial addresses and automorphism permutations based on precomputed powers of 5. Fig.~\ref{fig:perm}(b) shows an example of temporal permutation.

To support adaptive parallelism, we set the physical permutation width to 256 to reduce routing complexity and improve timing closure. Implementing a native 512-lane permutation network would require significant inter-SLR routing and deep pipelining, which would reduce the achievable frequency. Instead, during regular execution, we pack two 32-bit coefficients from different limbs into a single 64-bit word and permute two limbs simultaneously. This effectively provides a logical permutation throughput of 512 coefficients per cycle, while KLSS operations use the native 256-lane mode for 62-bit coefficients. In this way, the permutation unit supports both 256-way and 512-way parallelism, matching the adaptive parallelism of the compute array.

\subsection{Multi-bank Scratchpad}
The on-chip scratchpad buffers various limbs used during FHE execution, including Ct/Pt limbs, twiddle factors, and evaluation keys. We design a multi-bank scratchpad to simultaneously serve memory requests from the gather–scatter unit for HBM traffic and from the datapath scheduler for the compute array and permute unit.
The scratchpad is organized into eight dual-port banks, each with a width of 256 64-bit words. With this organization, each bank can serve either 256 62-bit coefficients or 512 32-bit coefficients per access, enabling support for the adaptive parallelism required by KLSS and regular execution.
Each bank is implemented using BRAM or URAM blocks through packing and stacking to achieve the required width and sufficient depth. The scratchpad capacity is sized to store the limbs required by the memory-efficient KLSS datapath, allowing the accelerator to avoid intermediate Ct transfers to off-chip memory.

\subsection{Datapath Scheduler}
The datapath scheduler is implemented as a state machine that orchestrates both computation and permutation tasks. It has two main functions.
First, the scheduler maintains Ct metadata at runtime and adaptively switches between KLSS and HKS based on the current Ct level and pre-determined switching points derived from the performance model.
The control sequences for both KLSS and HKS are precompiled as sequences of primitive operations, and the scheduler dynamically selects and invokes the appropriate sequence based on runtime decisions.
Second, the scheduler overlaps computation and permutation tasks under different parallelism modes. Since the permute unit has a physical throughput of 256 lanes, different scheduling strategies are required for KLSS and regular execution. For KLSS execution, the compute array operates with 256-way parallelism, which matches the permutation throughput. Therefore, while the permutation unit processes the current limb, the compute array processes the next limb, allowing permutation and computation to be fully overlapped during $\mathsf{(i)NTT}$.
For regular execution, the compute array operates with 512-way parallelism, which exceeds the physical throughput of the permute unit. To address this mismatch, we pack two 32-bit coefficients from two different limbs at the same coefficient index into one 64-bit word and permute two limbs simultaneously, as discussed in Sec.\,\ref{sec:perm}. 
This effectively provides a logical throughput of 512 coefficients per cycle.
With this packing strategy, permutation throughput matches compute throughput, enabling overlap between permutation and computation for regular execution as well.

Through adaptive scheduling and data packing, the scheduler ensures that permutation and computation can be overlapped in both KLSS and regular execution modes, maximizing hardware utilization and minimizing permutation overhead.

\subsection{Resource Utilization}
Table~\ref{tab:resource} summarizes the resource utilization of the proposed accelerator. DSP resources are consumed only by the compute array, with each compute unit using 34 DSPs. In addition to the scratchpad memory, BRAM resources are used to implement buffers for temporal permutation in the permute unit and AXI interfaces for HBM communication.
\begin{table}[t]
    \caption{Resource Consumption of the Proposed Accelerator}
    \label{tab:resource}
    \centering
    \vspace{-0.1cm}
    \begin{tabular}{c|ccccc}
        \toprule    
         & DSP & BRAM & URAM & LUT & FF \\
        \toprule
        Compute Array & 8,704 & 0 & 0 & 650k & 1,223k\\
        Permute Unit & 0 & 256 & 0 & 194k & 270k \\
        Scratchpad Memory & 0 & 3,072 & 800 & 29k & 40k \\
        Others & 0 & 384 & 0 & 235k & 497k\\
        \midrule
        \multirow{2}{*}{Total} & 8,704 & 3,712 & 800 & 1,108k & 2,030K\\
         & [96\%] & [92\%] & [83\%] & [85\%] & [79\%] \\
        \bottomrule
    \end{tabular}
    \vspace{-0.2cm}
\end{table}

\section{Evaluation}
\subsection{Experimental Setup}
We implement the proposed accelerator in Verilog and SystemVerilog on an Alveo U280 FPGA using Vivado and Vitis 2023.1 for synthesis, place-and-route, and host–kernel integration. The host communicates with the FPGA kernel through PCIe using the XRT runtime library. Input data for each benchmark are preloaded into HBM, and results are transferred back to the host only after kernel execution completes. To isolate accelerator performance, all reported latency measurements exclude PCIe transfer time.

Table~\ref{tab:parameter_sets} summarizes the FHE parameter sets used in our evaluation. We select two parameter sets with different scales to study the trade-off between HKS and KLSS under different parameter regimes. Set-1 is a small parameter set that does not support bootstrapping and is suitable for shallow circuits. Set-2 is a large and practical parameter set that supports bootstrapping and matches the parameter configuration ($\{N,\log Q,\log P, dnum\}$) used in prior FPGA-based works~\cite{agrawal2023fab,zhu2025dahe,xu2025fast,huang2025effact,yang2025ola,xu2026hera} for fair comparison. Both parameter sets provide 128-bit security and comply with the NIST security standard~\cite{Chen2017NIST}.

\begin{table}[t]
    \centering
    \caption{FHE Parameter Sets used for Evaluation}
    \label{tab:parameter_sets}
    \vspace{-0.2cm}
    \resizebox{\columnwidth}{!}{%
    \begin{tabular}{c|ccccccccccc}
    \toprule
         & $N$ & $L$ & $k$ & $\log q$ & $\log t$ & \textit{dnum} & $\gamma$ & $\alpha$ & $\alpha'$ & $\tilde{\beta}$ (max)\\
    \midrule
        \textbf{Set-1} & $2^{14}$ & 8 & 1 & 32 & 62 & 1 & 5 & 9 & 2 & 4\\
        \textbf{Set-2} & $2^{16}$ & 38 & 15 & 32 & 62 & 3 & 9 & 13 & 2 & 6\\
    \bottomrule
    \end{tabular}}
\end{table}

\subsection{Key-switching Performance}

\begin{figure}[t]
    \centering
    \label{fig:kskbreakdown}
    \includegraphics[width=\linewidth]{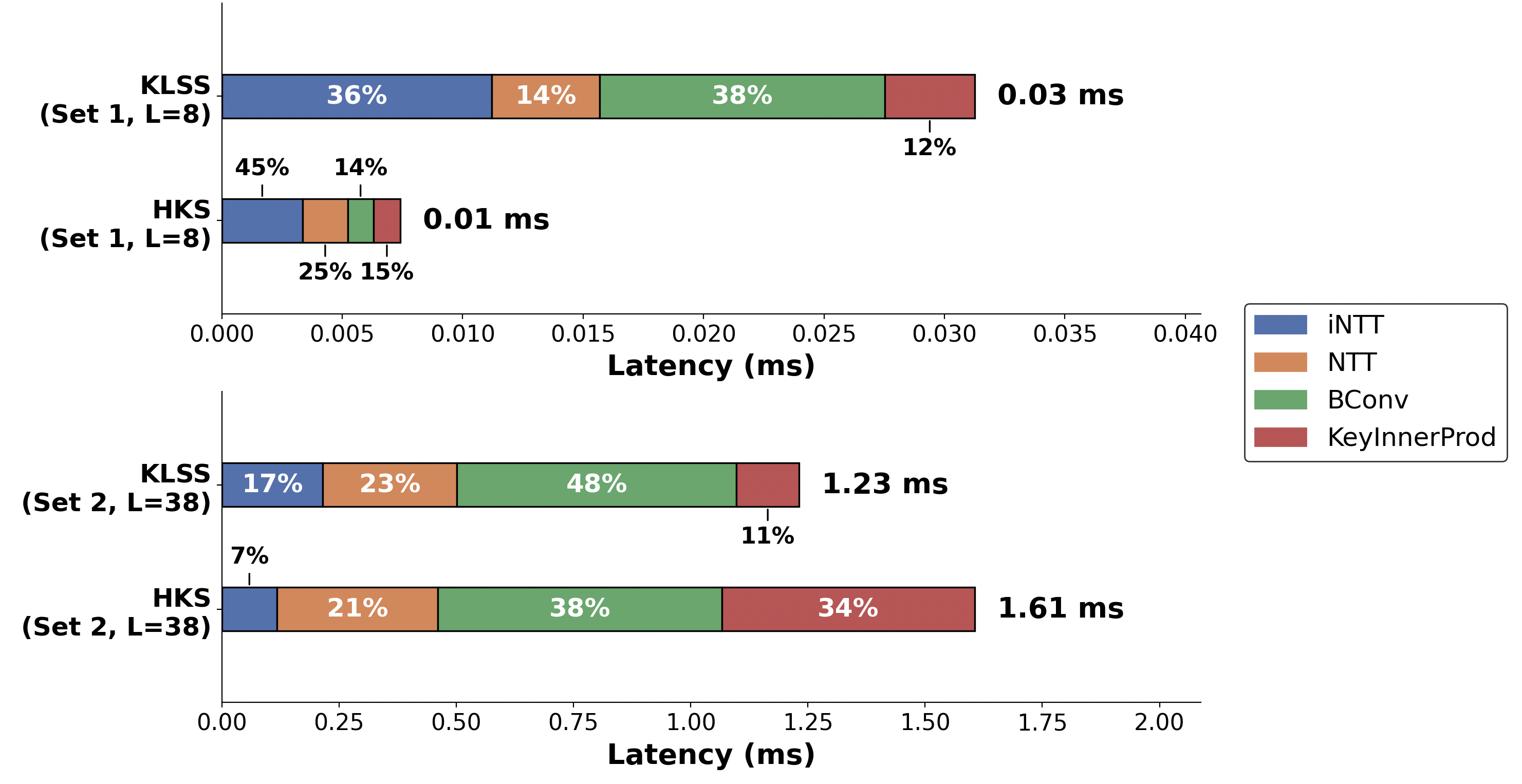} 
    \caption{HKS and KLSS performance breakdown and comparison}
\end{figure}

Fig.~5 shows the key-switching latency of KLSS and HKS on two parameter sets, along with the latency breakdown of major primitives. We observe that HKS and KLSS perform better on Set-1 and Set-2, respectively, which is consistent with our performance model and complexity analysis: HKS is more efficient at small Ct levels, while KLSS becomes more efficient at large Ct levels.

On Set-1, HKS is approximately $3\times$ faster than KLSS. This performance advantage mainly comes from the higher parallelism of HKS. In addition, HKS has lower complexity in certain primitives such as $\mathsf{BConv}$, which further reduces latency at small Ct levels where the total number of limbs is small.
On Set-2, where the Ct level is much larger, KLSS becomes more efficient and achieves about $1.3\times$ speedup over HKS. This is because KLSS uses 62-bit arithmetic in $\mathcal{R}_T$, which reduces the number of limbs involved in $\mathsf{KeyInnerProd}$ and $\mathsf{NTT}$, significantly lowering the computational complexity of these dominant operations. 

\subsection{Performance on FHE Operations}

\begin{table}[t]
  \centering
  \caption{Comparison of Basic Operations Against Prior Works in OP/s}
  \label{tab:latency_comparison}
  \resizebox{\columnwidth}{!}{%
  \begin{tabular}{l|cccc}
    \toprule
    \textbf{Work} & Device & Freq. (GHz) & $\mathsf{Mult}$ & $\mathsf{Rotate}$ \\
    \midrule
    \midrule
    100x (GPU)~\cite{jung2021over} & V100 & 1.2 & 337\,($2.26\times$) & 392\,($2.02\times$) \\
    FAB~\cite{agrawal2023fab} & U280 & 0.3 & 584\,($1.31\times$) & 636\,($1.25\times$) \\
    Poseidon~\cite{yang2023poseidon} & U280 & 0.45 & 273\,($2.79\times$) & 302\,($2.63\times$) \\
    DAHE~\cite{zhu2025dahe} & U280 & 0.3 & 653\,($1.17\times$) & 763\,($1.04\times$)\\
    OLA~\cite{yang2025ola} & U280 & 0.3 & 255\,($2.99\times$) & 266\,($2.98\times$) \\
    \textbf{This Work} & U280 & \textbf{0.3} & \textbf{763} & \textbf{793} \\
    \bottomrule
  \end{tabular}}
\end{table}

For basic operations, we benchmark $\mathsf{Mult}$ and $\mathsf{Rotate}$ and compare the results with prior work. These two operations are the dominant FHE primitives that require key-switching and are therefore significantly more expensive than simple operations such as $\mathsf{Add}$. Evaluating both operations on the large and practical parameter set allows us to demonstrate the benefits of adopting KLSS. Table\,\ref{tab:latency_comparison} summarizes the performance comparison with prior work in throughput (op/s).

On the large parameter set, our accelerator outperforms all prior FPGA-based accelerators for both operations. We achieve $1.17\times$--$2.99\times$ speedup for $\mathsf{Mult}$ and $1.04\times$--$2.98\times$ speedup for $\mathsf{Rotate}$. These results further demonstrate the advantage of KLSS in reducing key-switching complexity for large and practical parameter sets. Compared with a widely used GPU baseline, our accelerator achieves $2.26\times$ and $2.02\times$ speedups for $\mathsf{Mult}$ and $\mathsf{Rotate}$, respectively.

\subsection{Performance on Benchmarks}
Prior evaluations of key-switching and FHE operations are typically performed at the maximum Ct level ($L$) without considering the change in Ct levels during application execution. In this subsection, we evaluate our accelerator using application-level benchmarks where Ct levels decrease during execution. This allows us to evaluate the effectiveness of the proposed adaptive key-switching strategy. We evaluate our accelerator under three configurations based on the key-switching method: HKS-only, KLSS-only, and Adaptive. 

We use three widely-adopted benchmarks: (1) LoLA-MNIST~\cite{brutzkus2019low}: a small encrypted Convolutional Neural Network (CNN) for MNIST classification. LoLA-MNIST consumes only 5 Ct levels, and we use Set-1 for evaluation to align with prior work~\cite{fxhenn2023,yang2024framework,zhu2025dahe,yang2025ola} for fair comparison. (2) Bootstrapping, the key operation that enables deep FHE computation and end-to-end applications.  Bootstrapping requires a practical set, i.e., Set-2. We adopt the same bootstrapping algorithm as prior work~\cite{agrawal2023fab,huang2025effact,yang2025ola}, which consumes 15 Ct levels (i.e., $L_{boot}=15$). (3) Secure Cifar-10 image classification with ResNet-20 (RN-20)~\cite{lee2022privacy}. RN-20 is a deep benchmark that requires multiple rounds of bootstrapping, so we can fully evaluate the effectiveness of the adaptive solution by crossing the entire Ct level span multiple rounds.

Table~\ref{tab:mnist_comparison} shows the performance comparison of the small CNN benchmark against prior work. We observe that the HKS-only and Adaptive configurations achieve the same performance. This is because, for the small parameter set, HKS is always selected across all Ct levels (from level 8 to level 3) during execution, due to its advantage at small ciphertext levels. Overall, we achieve $1.14\times$--$58.5\times$ speedup over prior FPGA-based works. The performance improvement comes from higher compute parallelism compared to FxHENN~\cite{fxhenn2023} and Yang~\cite{yang2024framework}, and better compute unit utilization compared to DAHE~\cite{zhu2025dahe} and OLA~\cite{yang2025ola}, as our accelerator keeps the compute units busy throughout the benchmark execution.

\begin{table}[t]
  \centering
  \caption{Comparison of LoLa-MNIST CNN Inference Latency}
  \label{tab:mnist_comparison}
  \begin{tabular}{l|ccc}
    \toprule
    \textbf{Work} & Device & Freq. (GHz) & Latency (ms) \\
    \midrule
    \midrule
    FxHENN~\cite{fxhenn2023} & ACU15EG & - & 190 ($58.5\times$) \\
    Yang~\cite{yang2024framework} & U280 & 0.25 & 20 (6.15$\times$) \\
    DAHE~\cite{zhu2025dahe} & U280 & - & 3.7 ($1.14\times$) \\
    OLA~\cite{yang2025ola} & U280 & 0.3 & 10.0 ($3.08\times$)\\
    \midrule
    This Work (HKS-only) & U280 & 0.3 & 3.25 \\
    This Work (KLSS-only) & U280 & 0.3 & 14.63 \\
    \textbf{This Work (Adaptive)} & U280 & \textbf{0.3} & \textbf{3.25} \\
    \bottomrule
  \end{tabular}
\end{table}

Table~\ref{tab:boot_comparison} summarizes the results on deep benchmarks and compares latency with prior accelerators. For bootstrapping, the Adaptive configuration performs similarly to KLSS-only, which aligns with our expectations. Bootstrapping first restores the ciphertext level from 1 to $L=38$ using an inexpensive $\mathsf{ModUp}$ operation, after which the subsequent expensive operations consume $L_{boot}$ levels. As shown in Fig.~\ref{fig:motivation}, after bootstrapping, the remaining Ct level becomes $L_{eff}=23$, which is close to the switching boundary between HKS and KLSS. Therefore, most operations in bootstrapping occur at high Ct levels, where KLSS is more efficient, and the Adaptive configuration behaves similarly to KLSS-only.
Benefiting from the efficiency of KLSS at high Ct levels, our design achieves $1.84\times$--$3.31\times$ speedup over prior FPGA-based accelerators on bootstrapping. 
For the RN-20 benchmark, the Adaptive configuration achieves $2.08\times$ and $1.27\times$ speedups over the HKS-only and KLSS-only configurations, respectively. This is because deep benchmarks traverse the entire range of Ct levels multiple times, allowing the Adaptive scheme to select the more efficient key-switching method at every level. As a result, we observe $1.66\times$--$2.52\times$ speedup on RN-20 inference.
Overall, KLSS-only consistently outperforms HKS-only on large benchmarks because operations at high Ct levels dominate the total runtime, and KLSS is more efficient at high levels due to reduced algorithmic complexity.

\begin{table}[t!]
  \centering
  \caption{Comparison of deep benchmarks in Latency}
  \label{tab:boot_comparison}
  \resizebox{\columnwidth}{!}{%
  \begin{tabular}{l|cccc}
    \toprule
    \textbf{Work}  & Device & Freq. (GHz) & $T_{\text{boot}}$ (ms) & $T_{\text{RN-20}}$ (s) \\
    \midrule
    \midrule
    100x (GPU)~\cite{jung2021over}  & V100 & 1.2 & 328 & - \\
    FAB~\cite{agrawal2023fab}  & U280 & 0.3 & 92.4 & - \\
    Poseidon~\cite{yang2023poseidon} & U280 & 0.45 & - &2.67  \\
    EFFACT~\cite{huang2025effact} & U55C & 0.3 & - & 2.18 \\
    OLA~\cite{yang2025ola} & U280 & 0.3 & 166 & 3.30 \\
    \midrule
    This Work (HKS-only) & U280 & 0.3 & 91.9 & 2.72 \\
    This Work (KLSS-only) & U280 & 0.3 & 50.2 & 1.67 \\
    \textbf{This Work (Adaptive)} & U280 & \textbf{0.3} & \textbf{50.1} & \textbf{1.31} \\
    \bottomrule
  \end{tabular}}
\end{table}

\section{Related Work}



Hardware accelerators have been proposed to address the high computational and memory demands of FHE applications. Prior works~\cite{kim2022bts,samardzic2022craterlake,kim2022ark,kim2023sharp,agrawal2023fab,huang2025effact,xu2026hera,xu2024bandwidth, 11449120} primarily adopt the HKS key-switching method in their designs and evaluations. 
Recent works, namely FAST~\cite{fan2025fast} and HAWK~\cite{kong2025hawk}, were the first to integrate KLSS into ASIC accelerators and introduce adaptive key switching between HKS and KLSS. However, their designs rely heavily on large on-chip SRAM capacity, making them difficult to directly map onto memory-constrained FPGA platforms. In contrast, we propose a memory-efficient KLSS datapath tailored to FPGA platforms.
Jin \textit{et al.}~\cite{jin2025gpu} present the first GPU-based accelerator for KLSS. To the best of our knowledge, this work is the first FPGA-based accelerator to implement KLSS and evaluate adaptive key-switching on full FHE benchmarks.

\section{Conclusion}

In this paper, we presented a parallelism-adaptive FPGA accelerator for FHE that supports both HKS and KLSS key-switching methods. We showed that the efficiency of HKS and KLSS depends on the Ct level, with a trade-off between parallelism and algorithmic complexity. To exploit this observation, we developed a performance model to determine the optimal key-switching method enabling runtime switching between the two methods.
To support KLSS efficiently on FPGA platforms, we proposed a memory-efficient KLSS datapath that avoids intermediate Ct transfers to off-chip memory.
Implemented on an Alveo U280, we demonstrated that the adaptive design achieves high performance across both shallow benchmarks with small parameter sets and deep workloads with large parameter sets.
Overall, this work shows that the support of adaptive key-switching methods and hardware parallelism can improve the performance of FPGA-based FHE acceleration.

\section*{Acknowledgment}
This work is supported by the U.S. National Science Foundation (NSF) under grant CSSI-2311870. Jayashree Adivarahan is supported by the NSF Graduate Research Fellowship Program (GRFP). Equipment and support from AMD AECG are greatly appreciated. 

\textbf{Distribution Statement A:} Approved for public release. Distribution is unlimited.

\bibliographystyle{IEEEtran}
\bibliography{references}

@inproceedings{xu2025fast,
  title={FAST: FPGA Acceleration of Fully Homomorphic Encryption with Efficient Bootstrapping},
  author={Xu, Zhihan and Ye, Tian and Kannan, Rajgopal and Prasanna, Viktor K},
  booktitle={Proceedings of the 2025 ACM/SIGDA International Symposium on Field Programmable Gate Arrays},
  pages={115--126},
  year={2025}
}

@inproceedings{kong2025hawk,
  title={HAWK: Fully Homomorphic Encryption Accelerator with Fixed-Word Key Decomposition Switching},
  author={Kong, Liang and Fan, Shengyu and Deng, Xianglong and Chen, Lei and Fan, Guang and Shi, Guiming and Zhu, Yilan and Yang, Geng and Yan, Shoumeng and Zhang, Mingzhe},
  booktitle={Proceedings of the 58th Annual IEEE/ACM International Symposium on Microarchitecture (MICRO)},
  pages={1719--1734},
  year={2025}
}

@inproceedings{fan2025fast,
  title={FAST: An FHE Accelerator for Scalable-parallelism with Tunable-bit},
  author={Fan, Shengyu and Deng, Xianglong and Kong, Liang and Shi, Guiming and Fan, Guang and Meng, Dan and Hou, Rui and Zhang, Mingzhe},
  booktitle={Proceedings of the 52nd Annual International Symposium on Computer Architecture (ISCA)},
  pages={92--106},
  year={2025}
}

@inproceedings{kim2022ark,
  title={Ark: Fully homomorphic encryption accelerator with runtime data generation and inter-operation key reuse},
  author={Kim, Jongmin and Lee, Gwangho and Kim, Sangpyo and Sohn, Gina and Rhu, Minsoo and Kim, John and Ahn, Jung Ho},
  booktitle={Proceedings of the 55th IEEE/ACM International Symposium on Microarchitecture (MICRO)},
  pages={1237--1254},
  year={2022},
  organization={IEEE}
}

@inproceedings{kim2023sharp,
  title={SHARP: A short-word hierarchical accelerator for robust and practical fully homomorphic encryption},
  author={Kim, Jongmin and Kim, Sangpyo and Choi, Jaewan and Park, Jaiyoung and Kim, Donghwan and Ahn, Jung Ho},
  booktitle={Proceedings of the 50th Annual International Symposium on Computer Architecture (ISCA)},
  pages={1--15},
  year={2023}
}

@inproceedings{kim2022bts,
  title={Bts: An accelerator for bootstrappable fully homomorphic encryption},
  author={Kim, Sangpyo and Kim, Jongmin and Kim, Michael Jaemin and Jung, Wonkyung and Kim, John and Rhu, Minsoo and Ahn, Jung Ho},
  booktitle={Proceedings of the 49th Annual International Symposium on Computer Architecture (ISCA)},
  pages={711--725},
  year={2022}
}

@inproceedings{samardzic2022craterlake,
  title={Craterlake: a hardware accelerator for efficient unbounded computation on encrypted data},
  author={Samardzic, Nikola and Feldmann, Axel and Krastev, Aleksandar and Manohar, Nathan and Genise, Nicholas and Devadas, Srinivas and Eldefrawy, Karim and Peikert, Chris and Sanchez, Daniel},
  booktitle={Proceedings of the 49th Annual International Symposium on Computer Architecture (ISCA)},
  pages={173--187},
  year={2022}
}

@inproceedings{yang2023poseidon,
  title={Poseidon: Practical homomorphic encryption accelerator},
  author={Yang, Yinghao and Zhang, Huaizhi and Fan, Shengyu and Lu, Hang and Zhang, Mingzhe and Li, Xiaowei},
  booktitle={2023 IEEE International Symposium on High-Performance Computer Architecture (HPCA)},
  pages={870--881},
  year={2023},
  organization={IEEE}
}

@inproceedings{yang2023fpga,
  title={Fpga acceleration of rotation in homomorphic encryption using dynamic data layout},
  author={Yang, Yang and Long, Weihang and Kannan, Rajgopal and Prasanna, Viktor K},
  booktitle={2023 33rd International Conference on Field-Programmable Logic and Applications (FPL)},
  pages={174--181},
  year={2023},
  organization={IEEE}
}

@article{jung2021over,
  title={Over 100x faster bootstrapping in fully homomorphic encryption through memory-centric optimization with GPUs},
  author={Jung, Wonkyung and Kim, Sangpyo and Ahn, Jung Ho and Cheon, Jung Hee and Lee, Younho},
  journal={IACR Transactions on Cryptographic Hardware and Embedded Systems},
  pages={114--148},
  year={2021}
}

@article{lee2022privacy,
  title={Privacy-preserving machine learning with fully homomorphic encryption for deep neural network},
  author={Lee, Joon-Woo and Kang, HyungChul and Lee, Yongwoo and Choi, Woosuk and Eom, Jieun and Deryabin, Maxim and Lee, Eunsang and Lee, Junghyun and Yoo, Donghoon and Kim, Young-Sik and others},
  journal={iEEE Access},
  volume={10},
  pages={30039--30054},
  year={2022},
  publisher={IEEE}
}

@article{de2021does,
  title={Does fully homomorphic encryption need compute acceleration?},
  author={de Castro, Leo and Agrawal, Rashmi and Yazicigil, Rabia and Chandrakasan, Anantha and Vaikuntanathan, Vinod and Juvekar, Chiraag and Joshi, Ajay},
  journal={arXiv preprint arXiv:2112.06396},
  year={2021}
}

@inproceedings{agrawal2023fab,
  title={FAB: An FPGA-based accelerator for bootstrappable fully homomorphic encryption},
  author={Agrawal, Rashmi and de Castro, Leo and Yang, Guowei and Juvekar, Chiraag and Yazicigil, Rabia and Chandrakasan, Anantha and Vaikuntanathan, Vinod and Joshi, Ajay},
  booktitle={2023 IEEE International Symposium on High-Performance Computer Architecture (HPCA)},
  pages={882--895},
  year={2023},
  organization={IEEE}
}

@inproceedings{huang2025effact,
  title={EFFACT: A Highly Efficient Full-Stack FHE Acceleration Platform},
  author={Huang, Yi and Gong, Xinsheng and Kong, Xiangyu and Chen, Dibei and Zhu, Jianfeng and Zhu, Wenping and Li, Liangwei and Gao, Mingyu and Wei, Shaojun and Zhang, Aoyang and others},
  booktitle={2025 IEEE International Symposium on High Performance Computer Architecture (HPCA)},
  pages={1143--1157},
  year={2025},
  organization={IEEE}
}

@inproceedings{fullrns,
  title={A full RNS variant of approximate homomorphic encryption},
  author={Cheon, Jung Hee and Han, Kyoohyung and Kim, Andrey and Kim, Miran and Song, Yongsoo},
  booktitle={International Conference on Selected Areas in Cryptography},
  pages={347--368},
  year={2018},
  organization={Springer}
}

@inproceedings{yang2025ola,
  title={OLA: An FPGA-based Overlay Accelerator for Privacy Preserving Machine Learning with Homomorphic Encryption},
  author={Yang, Yang and Kannan, Rajgopal and Prasanna, Viktor K},
  booktitle={Proceedings of the 2025 ACM/SIGDA International Symposium on Field Programmable Gate Arrays},
  pages={127--138},
  year={2025}
}

@inproceedings{chen2015automatic,
  title={Automatic generation of high throughput energy efficient streaming architectures for arbitrary fixed permutations},
  author={Chen, Ren and Prasanna, Viktor K},
  booktitle={2015 25th International Conference on Field Programmable Logic and Applications (FPL)},
  pages={1--8},
  year={2015},
  organization={IEEE}
}

@inproceedings{cheon2017homomorphic,
  title={Homomorphic encryption for arithmetic of approximate numbers},
  author={Cheon, Jung Hee and Kim, Andrey and Kim, Miran and Song, Yongsoo},
  booktitle={International conference on the theory and application of cryptology and information security},
  pages={409--437},
  year={2017},
  organization={Springer}
}

@inproceedings{xu2024bandwidth,
  title={Bandwidth Efficient Homomorphic Encrypted Discrete Fourier Transform Acceleration on FPGA},
  author={Xu, Zhihan and Yang, Yang and Kannan, Rajgopal and Prasanna, Viktor K},
  booktitle={2024 IEEE 32nd Annual International Symposium on Field-Programmable Custom Computing Machines (FCCM)},
  pages={1--12},
  year={2024},
  organization={IEEE}
}

@inproceedings{fxhenn2023,
    author={Zhu, Yilan and Wang, Xinyao and Ju, Lei and Guo, Shanqing},
    title={{FxHENN}: {FPGA}-based acceleration framework for homomorphic encrypted {CNN} inference},
    booktitle={2023 IEEE International Symposium on High-Performance Computer Architecture (HPCA)},
    year={2023},
    pages={896--907},
    organization={IEEE}
}

@inproceedings{kim2019fpga,
  title={FPGA-based accelerators of fully pipelined modular multipliers for homomorphic encryption},
  author={Kim, Sunwoong and Lee, Keewoo and Cho, Wonhee and Cheon, Jung Hee and Rutenbar, Rob A},
  booktitle={2019 International Conference on ReConFigurable Computing and FPGAs (ReConFig)},
  pages={1--8},
  year={2019},
  organization={IEEE}
}

@inproceedings{singh2025ntt,
  title={NTT-SAA: Exploring NTT Acceleration with 2-D Systolic Array Architecture on FPGAs},
  author={Singh, Ashwajit and Xu, Zhihan and Prasanna, Viktor K},
  booktitle={2025 IEEE High Performance Extreme Computing Conference (HPEC)},
  pages={1--7},
  year={2025},
  organization={IEEE}
}

@article{zhu2025dahe,
    author={Zhu, Yilan and You, Honghui and Zhang, Wei and Xu, Jiming and Lou, Qian and Yan, Shoumeng and Ju, Lei},
    title={{DAHE}: Parameter-Adaptive and Memory Efficient {FPGA} Acceleration of Homomorphic Encryption},
    journal={IEEE Transactions on Computers},
    year={2025},
    publisher={IEEE},
    doi={10.1109/TC.2025.3569159}
}

@inproceedings{yang2024framework,
    author={Yang, Yang and Kannan, Rajgopal and Prasanna, Viktor K.},
    title={A Framework for Generating Accelerators for Homomorphic Encryption Operations on {FPGAs}},
    booktitle={2024 IEEE 35th International Conference on Application-specific Systems, Architectures and Processors (ASAP)},
    year={2024},
    pages={61--70},
    organization={IEEE}
}

@article{turan2020heaws,
  title={HEAWS: An accelerator for homomorphic encryption on the Amazon AWS FPGA},
  author={Turan, Furkan and Roy, Sujoy Sinha and Verbauwhede, Ingrid},
  journal={IEEE Transactions on Computers},
  volume={69},
  number={8},
  pages={1185--1196},
  year={2020},
  publisher={IEEE}
}

@inproceedings{lu2024fpga,
  title={An FPGA-based key-switching accelerator with ultra-high throughput for FHE},
  author={Lu, Zhaojun and Xu, Peng and Wang, Yijie and Yang, Yifan and Chen, Qidong and Yu, Weizong and Qu, Gang},
  booktitle={Proceedings of the 43rd IEEE/ACM International Conference on Computer-Aided Design},
  pages={1--9},
  year={2024}
}

@inproceedings{kumarathunga2025autontt,
  title={Autontt: Automatic architecture design and exploration for number theoretic transform acceleration on fpgas},
  author={Kumarathunga, Dilshan and Hu, Qilin and Fang, Zhenman},
  booktitle={2025 IEEE 33rd Annual International Symposium on Field-Programmable Custom Computing Machines (FCCM)},
  pages={1--9},
  year={2025},
  organization={IEEE}
}

@inproceedings{zhu2026efficient,
  title={An Efficient and Scalable Hardware Architecture for Number Theoretic Transform on FPGA with Design Automation},
  author={Zhu, Yilan and Yang, Geng and Tian, Xingyu and Kumarathunga, Dilshan and Kong, Liang and Deng, Xianglong and Fan, Shengyu and Fan, Guang and Shi, Guiming and Chen, Lei and others},
  booktitle={2026 IEEE International Symposium on High Performance Computer Architecture (HPCA)},
  pages={1--14},
  year={2026},
  organization={IEEE}
}

@inproceedings{xu2026hera,
  title={HERA: A Bandwidth-efficient Accelerator for Fully H omomorphic E nc r yption on HBM-enabled FPG A},
  author={Xu, Zhihan and Kannan, Rajgopal and Prasanna, Viktor},
  booktitle={Proceedings of the 2026 ACM/SIGDA International Symposium on Field Programmable Gate Arrays},
  pages={265--276},
  year={2026}
}

@inproceedings{neda2024ciflow,
  title={CiFlow: Dataflow analysis and optimization of key switching for homomorphic encryption},
  author={Neda, Negar and Ebel, Austin and Reynwar, Benedict and Reagen, Brandon},
  booktitle={2024 IEEE International Symposium on Performance Analysis of Systems and Software (ISPASS)},
  pages={61--72},
  year={2024},
  organization={IEEE}
}

@inproceedings{bajard2016full,
  title={A full RNS variant of FV like somewhat homomorphic encryption schemes},
  author={Bajard, Jean-Claude and Eynard, Julien and Hasan, M Anwar and Zucca, Vincent},
  booktitle={International Conference on Selected Areas in Cryptography},
  pages={423--442},
  year={2016},
  organization={Springer}
}

@inproceedings{halevi2019improved,
  title={An improved RNS variant of the BFV homomorphic encryption scheme},
  author={Halevi, Shai and Polyakov, Yuriy and Shoup, Victor},
  booktitle={Cryptographers’ Track at the RSA Conference},
  pages={83--105},
  year={2019},
  organization={Springer}
}

@inproceedings{kim2023accelerating,
  title={Accelerating HE operations from key decomposition technique},
  author={Kim, Miran and Lee, Dongwon and Seo, Jinyeong and Song, Yongsoo},
  booktitle={Annual International Cryptology Conference},
  pages={70--92},
  year={2023},
  organization={Springer}
}

@inproceedings{han2020better,
  title={Better bootstrapping for approximate homomorphic encryption},
  author={Han, Kyoohyung and Ki, Dohyeong},
  booktitle={Cryptographers’ Track at the RSA Conference},
  pages={364--390},
  year={2020},
  organization={Springer}
}

@article{liang2022number,
  title={Number theoretic transform and its applications in lattice-based cryptosystems: A survey},
  author={Liang, Zhichuang and Zhao, Yunlei},
  journal={arXiv preprint arXiv:2211.13546},
  year={2022}
}

@article{jin2025gpu,
  title={GPU Acceleration for KLSS Key Switching in Fully Homomorphic Encryption},
  author={Jin, Shutong and Cheung, Ray CC},
  journal={Mathematics},
  volume={13},
  number={23},
  pages={3809},
  year={2025},
  publisher={MDPI}
}

@book{solinas1999generalized,
  title={Generalized mersenne numbers},
  author={Solinas, Jerome A and others},
  year={1999},
  publisher={Faculty of Mathematics, University of Waterloo Waterloo, ON, Canada}
}

@techreport{Chen2017NIST,
  author      = {Hao Chen and Kim Laine and Rachel Player},
  title       = {Security of Homomorphic Encryption},
  institution = {National Institute of Standards and Technology (NIST) / Microsoft Research},
  year        = {2017},
  note        = {Presented at the NIST Workshop on White Box Cryptography},
}

@inproceedings{brutzkus2019low,
  title={Low latency privacy preserving inference},
  author={Brutzkus, Alon and Gilad-Bachrach, Ran and Elisha, Oren},
  booktitle={International conference on machine learning},
  pages={812--821},
  year={2019},
  organization={PMLR}
}

@inproceedings{11449120,
  author={Xu, Zhihan and Kannan, Rajgopal and Prasanna, Viktor K.},
  booktitle={2025 35th International Conference on Field-Programmable Logic and Applications (FPL)}, 
  title={FAME: FPGA Acceleration of Secure Matrix Multiplication with Homomorphic Encryption}, 
  year={2025},
  pages={100-109},
  doi={10.1109/FPL68686.2025.00025}}

\end{document}